# Self-Assembly of and Plasmon-Enhanced Ultrafast Magnetization in Ag–Co Hybrid Nanoparticles

*Katsura Ikemiyas, Kuniaki Konishi, Eiko Fujii,*
*Toshihiro Kogure, Makoto Kuwata-Gonokami\*, Tetsuya Hasegawa*

Katsura Ikemiya, Kuniaki Konishi, Eiko Fujii, Prof. Toshihiro Kogure, Prof. Makoto Kuwata-Gonokami, Prof. Tetsuya Hasegawa
The University of Tokyo, Bunkyo-ku, Tokyo 113-0033, Japan
E-mail: gonokami@phys.s.u-tokyo.ac.jp

Ultrafast demagnetization in magnetic nanoparticles using pulsed laser has attracted considerable attention because of its potential applications in spintronics, such as data storage. In such applications, it is necessary to control magnetization using low-energy laser pulses; however, this poses the problem of increasing the amount of energy from the excitation laser pulses to the spin subsystem. We take advantage of the phenomenon known as localized surface plasmon resonance (LSPR) to enhance the energy transfer from laser pulses to the spin subsystem. To induce LSPR, hybrid nanoparticles consisting of noble metal nanoparticles with LSPR absorption and magnetic metal nanoparticles are prepared using a novel method. Specifically, Ag–Co hybrid nanoparticles are prepared by a self-assembly method using pulsed laser deposition. We performed measurements of the static Faraday and time-resolved Faraday effects using a pump–probe technique on the Ag–Co hybrid nanoparticles with various Ag–Co ratios. The data suggest that the LSPR absorption and demagnetization amplitude increase with the increasing Ag–Co ratio. The results indicate that the amount of energy transferred from the laser pulses to the spin system of magnetic nanoparticles can increase via LSPR absorption.

# 1. Introduction

Ultrafast demagnetization, which is the reduction of magnetization within a few picoseconds after femtosecond laser excitation, has attracted considerable attention as a technique for the ultrafast manipulation of magnetization. Laser-induced ultrafast demagnetization was first observed in Ni thin films[1] and has since been observed in nanoscale magnetic materials, such as thin-film structures[1–3] or nanoparticles.[4,5,6,7] In spintronics[8], such as data storage[9], it is necessary to control the ultrafast magnetization of magnetic nanoparticles using low-energy laser pulses; however, this poses the problem of increasing the amount of energy transferred from the excitation laser pulses to the spin subsystem.

To overcome this problem, we took advantage of a phenomenon known as localized surface plasmon resonance (LSPR). LSPR is the collective oscillation of free electrons in metal nanoparticles, which is induced when the free electrons are coupled with electromagnetic waves of light at resonant frequency. When the frequency of incident light approaches that of LSPR, the amplitude of the electromagnetic waves near the nanoparticles drastically increases. Consequently, the cross-section of linear and nonlinear interactions, such as absorption, fluorescence, the Raman effect, and second- and third-harmonic generation, drastically increase.[10,11] The higher the excitation pulse energy, the larger the ultrafast demagnetization.[5,12] Thus, LSPR is expected to enhance the ultrafast demagnetization, because it increases the amount of energy from the excitation laser pulses to the spin subsystem. However, this has not been experimentally demonstrated to date.

Superparamagnetic nanoparticles are the most suitable magnetic materials for observing ultrafast demagnetization, because they have the lowest magnetic anisotropy.[4] However, unlike noble metal nanoparticles, magnetic metal nanoparticles do not exhibit strong LSPR. Based on this fact, we prepared and used hybrid nanostructures consisting of noble metal and ferromagnetic metal nanoparticles. In such hybrid nanoparticles, the electric field strongly increases near the noble metal nanoparticles via LSPR. Thus, the enhanced field is expected to couple with the nearby ferromagnetic metal nanoparticles, as shown in **Figure 1**.

Both superparamagnetism and LSPR require particles as small as several nanometers. In addition, the hybrid nanoparticles are preferably oriented because LSPR is sensitive to the shape and orientation of the noble metal nanoparticles. To date, most hybrid nanoparticles have been prepared by liquid-route techniques, such as decomposition of metallic salts.[13–20] The nanoparticles obtained by such methods are not oriented; moreover, they are randomly dispersed in the liquid. To obtain oriented fine structures, lithographic techniques have been used;[21–26] however, there is room for improvement. For example, photolithograpy has high throughput, but it requires a complicated multistep process. Electron-beam lithography produces fine structures of several nanometers, but it is time-consuming. Thus, a simple and expedient method for producing small and oriented hybrid nanoparticles is required. In this study, we developed and used a novel self-assembly method using pulsed laser deposition (PLD) to fabricate Ag–Co hybrid nanoparticles dispersed and aligned in a thin film. We chose $TiO_2$ as the matrix material, because $TiO_2$ is transparent in the wide wavelength range, Ti is immiscible with Ag and Co, and $TiO_2$ does not oxidize Ag or Co because Ti has smaller Gibbs free energy for oxidation. The obtained Co nanoparticles were small enough to show superparamagnetism and were adjoined to Ag nanoparticles. The hybrid nanoparticles were oriented in the matrix. To date, ultrafast demagnetization measurements have never been performed on such hybrid nanoparticles.

To examine the demagnetization enhancement via LSPR, we measured the static Faraday effect and time-resolved Faraday effect using a pump–probe technique. LSPR absorption and demagnetization amplitude were observed to increase with increasing Ag–Co ratio. The observations suggest that the demagnetization amplitude can be increased by increasing the LSPR absorption.

## 2. Experimental Section

The epitaxial thin films of (001)-oriented anatase $TiO_2$ containing Ag–Co hybrid nanoparticles were prepared on $LaSrAlO_4$ (LSAO) (001) single-crystal substrates using PLD. For ablation, a Kr–F excimer laser (wavelength = 248 nm) was operated with the laser fluence of 2 J cm$^{-2}$ per pulse and the repetition rate of 2 Hz.

The sintered pellets of pure $TiO_2$ and the mixtures of $TiO_2$, CoO, and $Ag_2O$ (molar Ti:Co:Ag of 95:5:$x$, where $x$ = 0, 5, 10, and 20) were used as PLD targets for the anatase $TiO_2$ seed layers and ($Ag_x$, Co):$TiO_2$ top layers containing Co and Ag, respectively. First, an anatase $TiO_2$ seed layer was deposited on the LSAO substrate at the substrate temperature ($T_s$) of 650 °C and oxygen pressure ($P_{O2}$) of 5 × 10$^{-3}$ Torr. Then, the ($Ag_x$, Co):$TiO_2$ top layer ($x$ = 0, 5, 10, and 20) was grown at $T_s$ = 300 °C and $P_{O2}$ = 1.0 × 10$^{-6}$ Torr. The prepared films are hereafter referred to as ($Ag_x$, Co):$TiO_2$/$TiO_2$ ($x$ = 0, 5, 10, and 20). The thicknesses of the seed layers and top layers were 5 nm and 28 nm for $x$ = 0 and 4, respectively, while the thicknesses of top layers were 24 nm for $x$ = 5, 10, and 20. Then, we measured the optical properties of the synthesized samples. In addition, a thicker film (seed layer = 6 nm, top layer = 36 nm) for $x$ = 20 was prepared to investigate the growth mechanisms of the Ag–Co hybrid nanoparticles.

The crystallinity and crystallographic orientation of the prepared films were evaluated by X-ray diffraction (XRD). The size and distribution of the Ag–Co hybrid nanoparticles were examined by transmission electron microscopy (TEM) and scanning TEM (STEM) equipped with an energy-dispersive X-ray detector (STEM–EDX) or with high-angle annular dark-field imaging (STEM–HAADF). The magneto-optical properties were measured using a magneto-optical spectrometer (BH-M800UV-KC-KF; Neoark Corp., Tokyo, Japan). Ultrafast demagnetization dynamics were established by measuring the time-resolved Faraday effect using a pump–probe technique. A regenerative amplified Ti:sapphire laser system (RegA9000, Coherent Inc.) operating at 120 kHz was used. The fundamental wavelength was 800 nm and the pulse duration was 220 fs. The wavelength of the probe pulses was set to 800 nm and that of the pump pulses, which was generated by frequency doubling, was 400 nm. The pump and probe beam diameters were approximately 0.2 mm. An external magnetic field of 9 kOe was applied perpendicular to the film surfaces. All measurements were performed at room temperature.

## 3. Results and Discussion

The cross-sectional and planar STEM–EDX and STEM–HADDF images of ($Ag_x$, Co):$TiO_2$/$TiO_2$ ($x$ = 0[27], 5, 10, and 20) films, shown in **Figure 2**(a–j), confirmed that the films consisted of Ag–Co hybrid nanoparticles embedded in the anatase $TiO_2$ matrix. As seen in the cross-sectional images (Figures 2(b, d, f, h, and j)), the Co nanoparticles formed at the interface between the seed and top layers, whereas the Ag nanoparticles formed on top of the Co nanoparticles, as seen in Figures 2(c–j). The comparison between the thinner and thicker ($Ag_{20}$, Co):$TiO_2$/$TiO_2$ films in Figures 2(g–j) disclosed that even though both films had Ag–Co hybrid nanoparticles consisting of Co nanospheres and Ag nanorods with almost the same number density (0.0057 $nm^{-2}$ for the thinner and 0.0084 $nm^{-2}$ for the thicker), the Co nanospheres were larger and the Ag nanorods were longer in the thicker film. This suggests that the Co atoms have stronger chemical affinity for the $TiO_2$ seed layer than the Ag atoms, the nuclei of the Ag–Co hybrid nanoparticles form at the beginning of the film growth, and the Co and Ag atoms are always phase-separated and never mix. The images in Figure 2(a–j) show that the Co nanoparticles are cylindrical for films with $x$ = 0 and 5, and spherical for films with $x$ = 10 and 20. Moreover, the size of the Co nanoparticles decreased with the increasing Ag content ($x$). In the films, the average volume of the Co nanoparticles was 1900 $nm^3$ (10 nm in diameter and 22 nm in height) for $x$ = 0, 450 $nm^3$ (6.7 nm in diameter and 13 nm in height) for $x$ = 5, 91 $nm^3$ (5.6 nm in diameter) for $x$ = 10, and 45 $nm^3$ (4.4 nm in diameter) for $x$ = 20. The results strongly suggest that the content of Ag atoms affects the diffusion and aggregation rate of Co atoms, resulting in changes in the nucleation density of the Co nanoparticles. The amount of Ag adjacent to Co nanoparticles increased with $x$ (Figure 2(a–j)).

**Figure 3**(a) shows a high-resolution TEM image of the thicker $(Ag_{20}, Co):TiO_2/TiO_2$ film. The $TiO_2$ matrix and Ag nanoparticles appear as bright and dark regions, respectively. Lattice fringes of 0.19 nm spacing corresponded to $d_{200}$ of anatase $TiO_2$, those of 0.20 nm to $d_{200}$ of fcc-Ag, and those of 0.24 nm to $d_{111}$ of fcc-Ag. No TEM lattice images of Ag nanoparticles or Co nanoparticles for the $(Ag_x, Co):TiO_2/TiO_2$ ($x$ = 5, 10, and 20) films were observed, because the nanoparticles were very small and at the boundary of the anatase $TiO_2$ seed layer and $(Ag_x, Co):TiO_2$ top layer. In the $(Ag_0, Co):TiO_2/TiO_2$ film[27], the Co nanoparticles have the fcc structure, which is the most stable structure for small Co nanoparticles. Thus, we speculate that the Co nanoparticles in the $(Ag_x, Co):TiO_2/TiO_2$ films ($x$ = 5, 10, and 20) possess the fcc structure. The XRD patterns of $(Ag_x, Co):TiO_2/TiO_2$ ($x$ = 0, 5, 10, and 20) in Figure 3(b) indicate that the anatase $TiO_2$ films grew epitaxially on the LSAO substrate with the (001) plane of anatase parallel to the (001) plane of LSAO. No diffraction peaks from fcc-Ag or fcc-Co were detected in the XRD patterns. This is probably because Ag and Co were randomly oriented and their concentration was too small.

**Figures 4**(a–c) depict the Faraday ellipticity spectra, the Faraday ellipticity vs magnetic field curves measured at 800 nm, and the absorption spectra of the films, respectively. The Faraday ellipticity spectra of $(Ag_x, Co):TiO_2/TiO_2$ ($x$ = 0, 5, 10, and 20) films were similar in shape and amplitude to one another, indicating that the amount of Co incorporated in each film was nearly identical. The $(Ag_0, Co):TiO_2/TiO_2$ film showed a hysteresis loop with coercivity value of ~1 kOe. In contrast, the $(Ag_x, Co):TiO_2/TiO_2$ ($x$ = 5, 10, and 20) films were superparamagnetic with small coercivity (~0.05 kOe or less). This is attributable to the reduction in the size of the Co nanoparticles with increasing Ag. In the absorption spectra, peaks at ~450 nm were observed for the $(Ag_x, Co):TiO_2/TiO_2$ ($x$ = 10 and 20) films, which indicate LSPR in the Ag nanoparticles. The red-shifting of the peak wavelength with increasing $x$ was also observed, suggesting that the Ag nanoparticles increased in size with $x$, as is typically seen with metal nanoparticles.[28] The peak intensity increased with increasing amount of Ag. The strong absorption in the shorter wavelength region (<350 nm) originated from the TiO$_2$ matrix. To excite LSPR, the pump pulse wavelength was set to 400 nm in the time-resolved measurements.

**Figure 5** shows the raw time-resolved differential Faraday ellipticity data for the (Ag$_{20}$, Co):TiO$_2$/TiO$_2$ film under polarities opposite to those of the external magnetic field ($\Delta\eta$ (+9 kOe): blue line and $\Delta\eta$ (−9 kOe): orange line) and without the external magnetic field ($\Delta\eta$ (0 kOe): green line). The polarity of the Faraday ellipticity depends on the direction of the external magnetic field. To find the magnetic and nonmagnetic components, the formulae $\Delta\eta = (\Delta\eta$ (+9 kOe) $- \Delta\eta$ (−9 kOe))/2 and $\Delta\eta_{nonmag} = (\Delta\eta$ (+9 kOe) $+ \Delta\eta$ (−9 kOe))/2 were used, respectively. As shown in Figure 5, the nonmagnetic component ($\Delta\eta_{nonmag}$: black line) was exactly the same as $\Delta\eta$ (0 kOe).

**Figure 6**(a) shows the differential Faraday ellipticity normalized by the saturation value of the static Faraday ellipticity ($\Delta\eta/\eta$) of the (Ag$_x$, Co):TiO$_2$/TiO$_2$ films measured under a pump laser fluence of 0.06 mJ cm$^{-2}$. The (Ag$_x$, Co):TiO$_2$/TiO$_2$ films exhibited ultrafast demagnetization over timescales of subpicoseconds and magnetization recovery over timescales of several tens of picoseconds. The observed demagnetization times were similar to those reported for Fe$_3$O$_4$[5] and Co nanoparticles[4], whereas the observed magnetization recovery times were much shorter than those reported for Fe$_3$O$_4$, Co$_x$Fe$_{3-x}$O$_4$, and Co[4,5,6,7] (several hundreds of picoseconds). The reason for the shorter recovery times in this study is probably that the pump laser energy was lower (0.06 mJ cm$^{-2}$) than that used for Fe$_3$O$_4$, Co$_x$Fe$_{3-x}$O$_4$, and Co (>1 mJ cm$^{-2}$).[4,5,6,7] Discussing the ultrafast magnetization mechanism is beyond the scope of this study. Thus, we focus on how the LSPR affects the ultrafast magnetization.

The peak demagnetization amplitude strongly depended on the Ag–Co ratio for $x$ = 5 to 20, whereas it did not depend for $x$ = 0 to 5, as shown in Figure 6(b), even though the saturation value of the static Faraday ellipticity was independent of $x$, as shown in Figure 4(a). Two factors are presumably responsible for the enhanced demagnetization: the decrease in the size of the Co nanoparticles with increasing $x$ and the increase in the LSPR absorption intensity of the Ag nanoparticles with increasing $x$. However, the contribution of the former is probably small. In Figure 6(b), the demagnetization amplitude, the normalized volume of each Co nanoparticle, and the LSPR absorption intensity at 400 nm are plotted against $x$. The demagnetization amplitude varied widely for $x$ = 5 to 20, whereas it hardly varied for $x$ = 0 to 5. This behavior is similar to that of the LSPR absorption intensity but differs from that of the volume of each Co nanoparticle.

The results confirm that LSPR enhances the demagnetization amplitude of the nanoparticles. Ultrafast demagnetization is considered the result of the thermalization of photoexcited hot electrons.[4] In general, the photoexcitation of the electronic subsystem of sparsely distributed ferromagnetic metal nanoparticles is inefficient. However, the absorption cross-section of the as-synthesized nanostructures, consisting of noble metal and ferromagnetic nanoparticles, increased via LSPR, leading to the efficient coupling of light with the electron subsystem of the Co nanoparticles.[29] Figure 6(c) shows the demagnetization amplitude of ($Ag_x$, Co):$TiO_2$/$TiO_2$ ($x$ = 0 and 20) films as a function of the pump laser fluence. In the applied range of pump laser fluence, we did not observe saturation of the demagnetization signal, whereas we clearly observed enhancement in the demagnetization signal resulting from the existence of Ag nanoparticles.

In all films, the magnetization recovery process was effectively expressed by combining fast ($\tau \approx$ 2 ps) and slow ($\tau \approx$ 10 ps) exponential decay components. Figure 6(a) shows the experimental data (colored lines) and fitting curves (black lines). **Figure 7**(a) shows the fitting parameters for the ($Ag_x$, Co):$TiO_2$/$TiO_2$ films as a function of $x$. As shown in the figure, the amplitude of the slower recovering component was almost the same in all films. However, the amplitude of the faster recovering component increased for $x$ = 5 to 20. The behavior of the amplitude of the faster recovering component is similar to that observed for the demagnetization amplitude. This confirms that the acceleration of the relaxation process was caused by the Ag nanoparticles near the Co nanoparticles. Such fast recovery has never been reported for single magnetic nanoparticles. The demagnetization of ($Ag_x$, Co):$TiO_2$/$TiO_2$ ($x$ = 0 and 20) films was measured at various pump laser fluences. The results are shown in **Figure 8**(a and b), and the fitted values of the fast- and slow-recovering components are summarized in Figure 7(b). As the pump laser fluence increased, only the amplitude of the slower recovering component increased, as in $Fe_3O_4$ nanoparticles.[5] The amplitude of the faster recovering component was almost constant.

Based on the above results, we propose the following demagnetization mechanism for the Ag–Co hybrid nanoparticles. The pump laser energy is absorbed by the Ag–Co hybrid nanoparticles and the $TiO_2$ matrix. Thus, the temperature in the Ag–Co hybrid nanoparticles increases because of the pump laser absorption by the Ag–Co hybrid nanoparticles and the heat transfer from the heated $TiO_2$ matrix. When the Ag–Co hybrid nanoparticles are irradiated with the pump laser pulse, strong LSPR is induced at the interface between the Co and Ag nanoparticles as they are intensively heated, as shown in Figure 1. Large amounts of heat are transferred from the pump laser pulse to the Ag–Co hybrid nanoparticles with increasing $x$, because the LSPR absorption intensity increases. The local heating of the Ag–Co hybrid nanoparticles relaxes quickly; therefore, the demagnetization amplitude and faster recovery component increase with $x$. The slower recovery component probably results from the cooling of the heated $TiO_2$ matrix.

## 4. Conclusion

In this study, we found that ultrafast demagnetization could be enhanced by taking advantage of the LSPR phenomenon even for low laser pulse energies. Suitably prepared hybrid nanoparticles significantly contributed to such behavior. The as-synthesized Ag–Co hybrid nanoparticles exhibited superparamagnetic properties and produced LSPR, which enabled us to examine how LSPR affected the ultrafast demagnetization. The results of this study can help to better understand and design new strategies for controlling ultrafast magnetization in nanostructures.

**Acknowledgements**
We would like to thank N. Kanda for performing the ultrafast magnetization measurements. This research is supported by the Photon Frontier Network Program, KAKENHI (20104002), Special Coordination Funds for Promoting Science and Technology of MEXT, Japan, and by JSPS through its FIRST Program.

Received: ((will be filled in by the editorial staff))
Revised: ((will be filled in by the editorial staff))
Published online: ((will be filled in by the editorial staff))

FIGURES

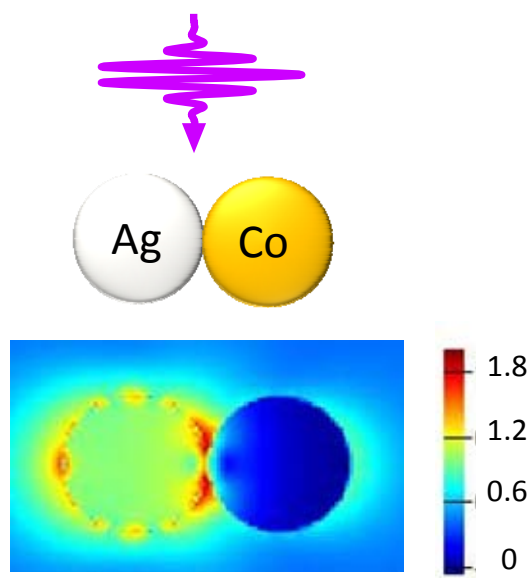

**Figure 1.** Normalized distribution of the calculated electric field intensity on the Ag–Co hybrid nanoparticles. The scale is logarithmic. The size of the nanoparticles is 10 nm and the incident wavelength for the electric field calculation is 365 nm. The calculations were performed using commercial finite-difference time-domain simulation software (FDTD solutions, Lumerical).

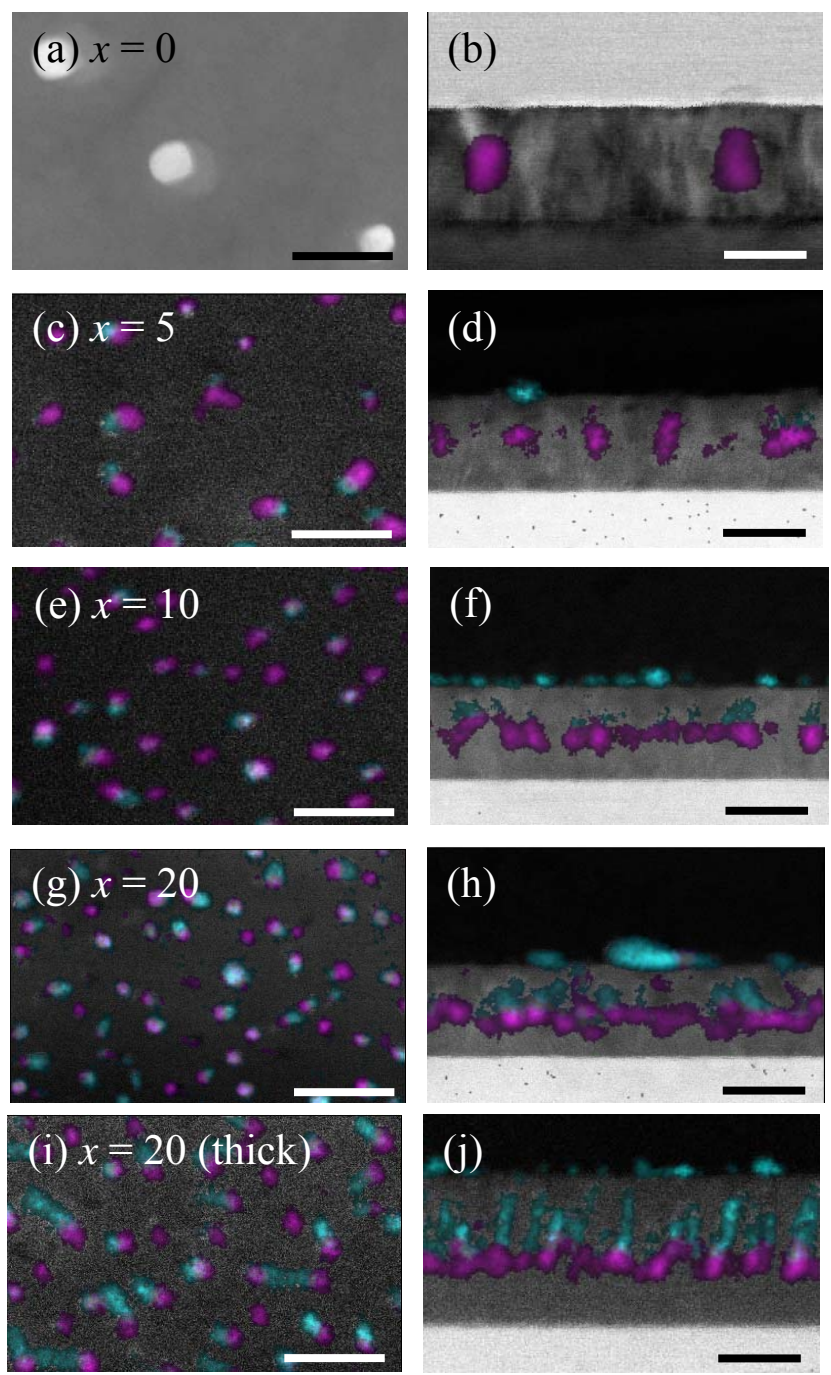

**Figure 2.** (a) Planar-view STEM–HAADF images of (Ag$_0$, Co):TiO$_2$/TiO$_2$ film and planar-view STEM–EDX images of Ag$_x$,Co:TiO$_2$/TiO$_2$ films where (c) $x$ = 5, (e) $x$ = 10, (g) $x$ = 20, and (i) $x$ = 20 (thick) films. Cross-sectional STEM–EDX images of (Ag$_x$, Co):TiO$_2$/TiO$_2$ films for (b) $x$ = 0, (d) $x$ = 5, (f) $x$ = 10, (h) $x$ = 20, and (j) $x$ = 20 (thick) films. The STEM–HAADF images show regions of Co (brighter parts) and the STEM–EDX images show regions of Co (purple) and Ag (blue). The length of all scale bars is 25 nm.

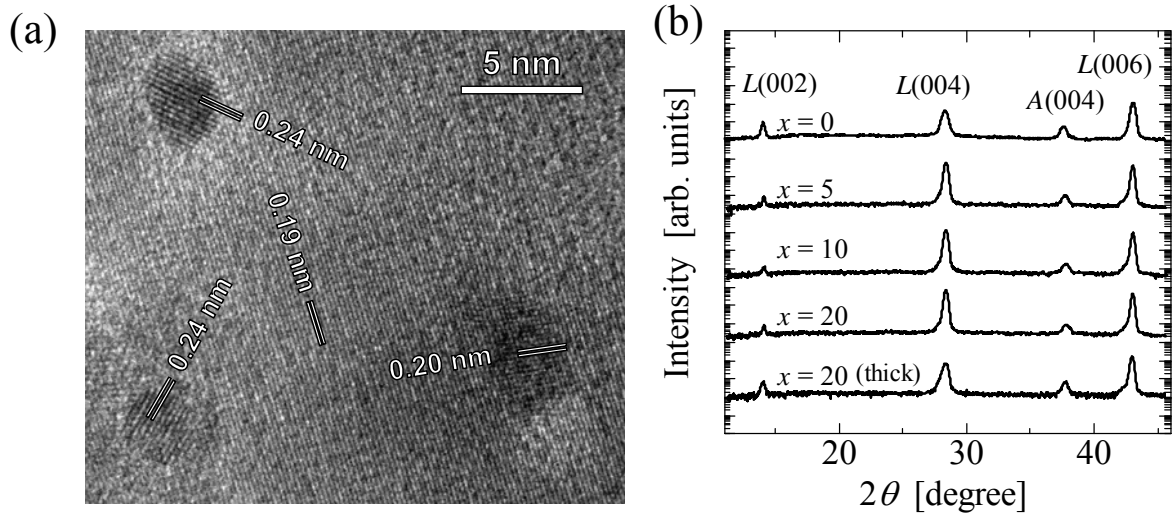

**Figure 3.** (a) Planar-view high-resolution TEM image of Ag nanoparticles in $(Ag_{20}, Co):TiO_2/TiO_2$ film. (b) XRD patterns of $(Ag_x, Co):TiO_2/TiO_2$ films. "A" and "L" denote the diffraction peaks of anatase $TiO_2$ and $LaSrAlO_4$, respectively.

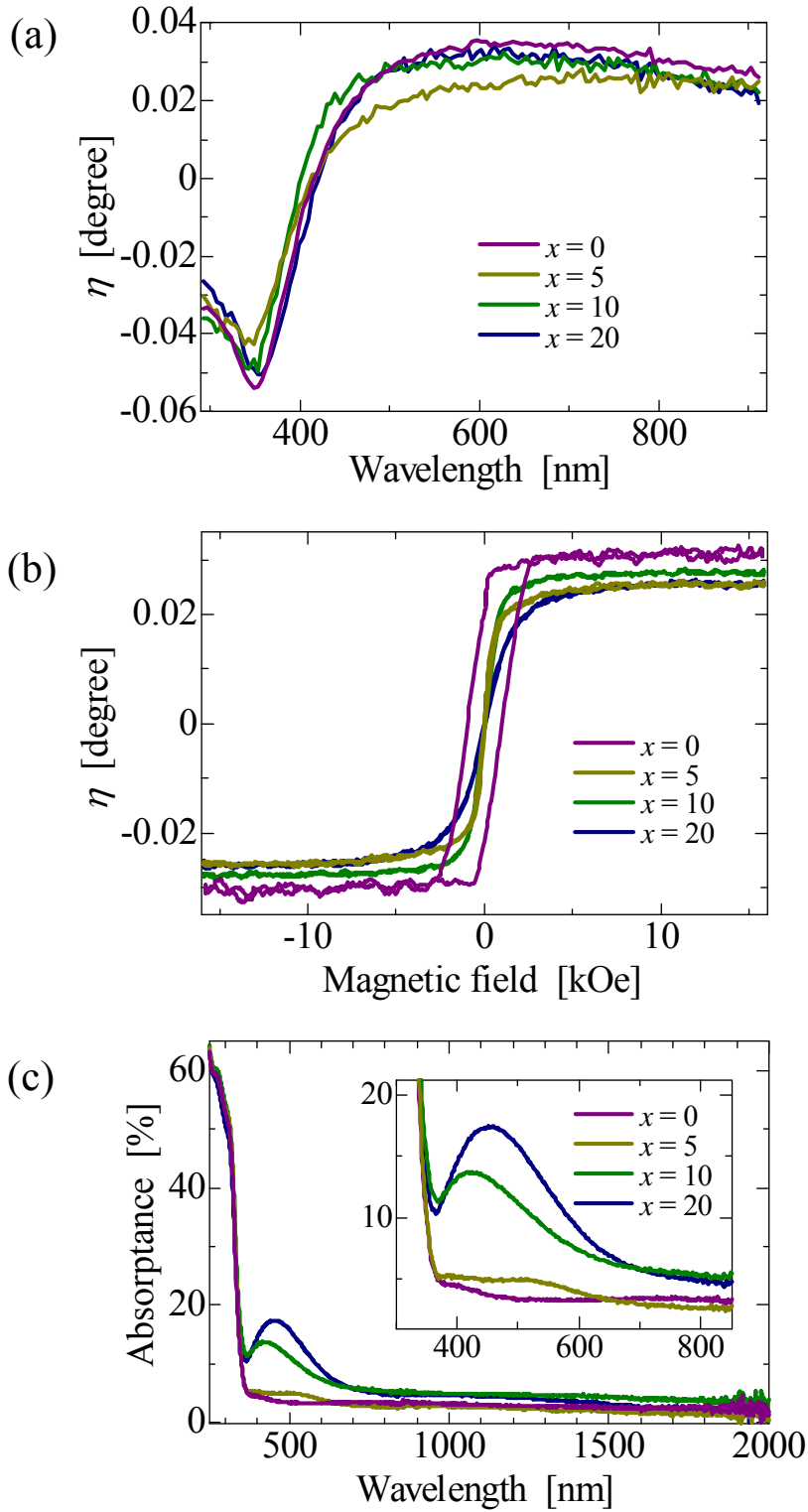

**Figure 4.** (a) Faraday ellipticity spectra, (b) Faraday ellipticity vs magnetic field curves measured at 800 nm, and (c) absorption spectra for $(Ag_x, Co):TiO_2/TiO_2$ films.

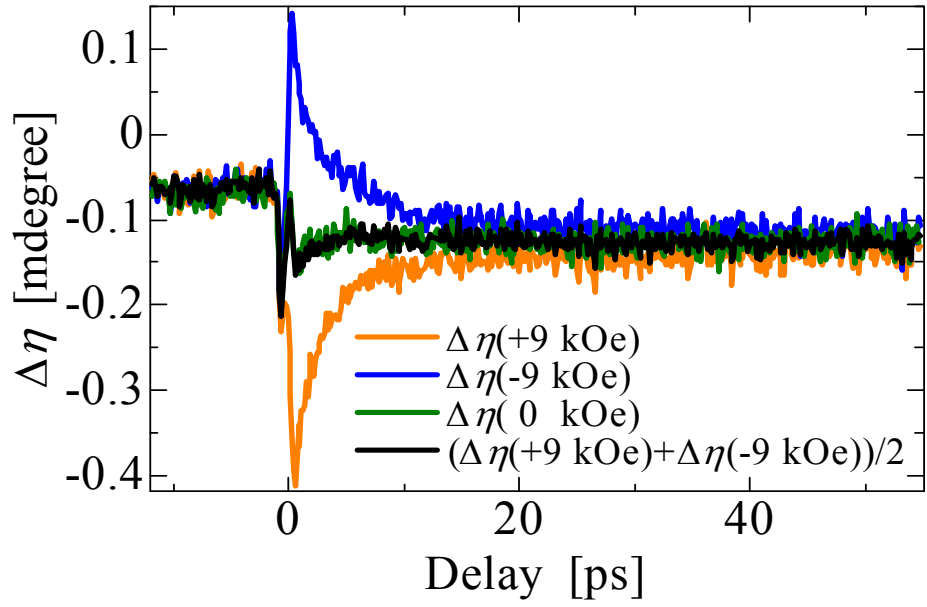

**Figure 5.** Raw time-resolved differential Faraday ellipticity data for the (Ag$_{20}$, Co):TiO$_2$/TiO$_2$ film under an external magnetic field of +9 kOe (orange), −9 kOe (blue), and 0 kOe (green). The black line represents the sum of the blue and orange curves.

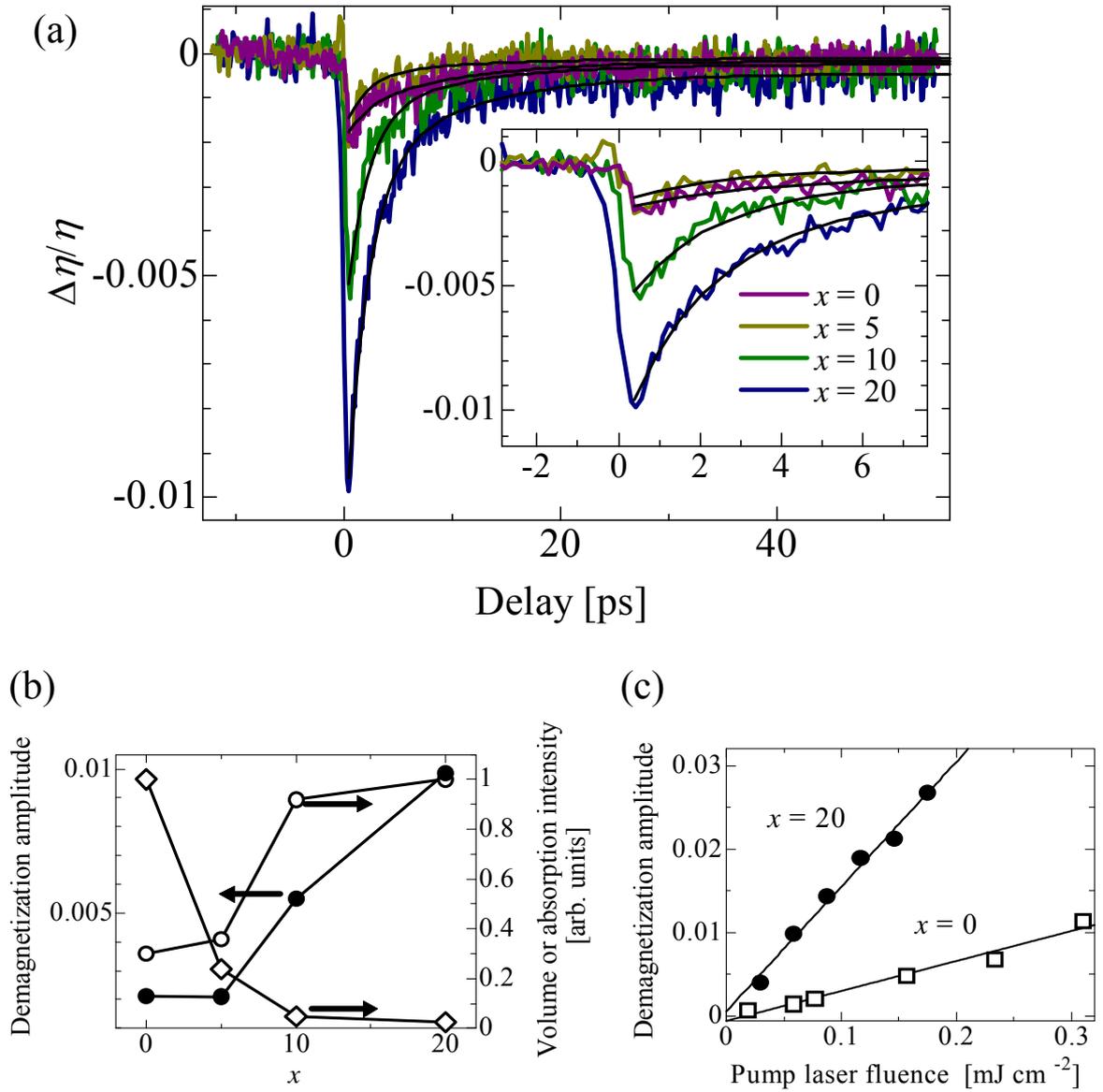

**Figure 6.** (a) $\Delta\eta/\eta$ of the $(Ag_x, Co):TiO_2/TiO_2$ films measured under a pump laser fluence of 0.06 mJ cm$^{-2}$. The black lines are fitted curves (see text for details). (b) Demagnetization amplitude (closed circles), normalized Co nanoparticle volume (open diamonds), and absorption intensity at 400 nm of the $(Ag_x, Co):TiO_2/TiO_2$ films (open circle). (c) Demagnetization amplitude of $(Ag_x, Co):TiO_2/TiO_2$ films for $x = 0$ (circles) and $x = 20$ (squares).

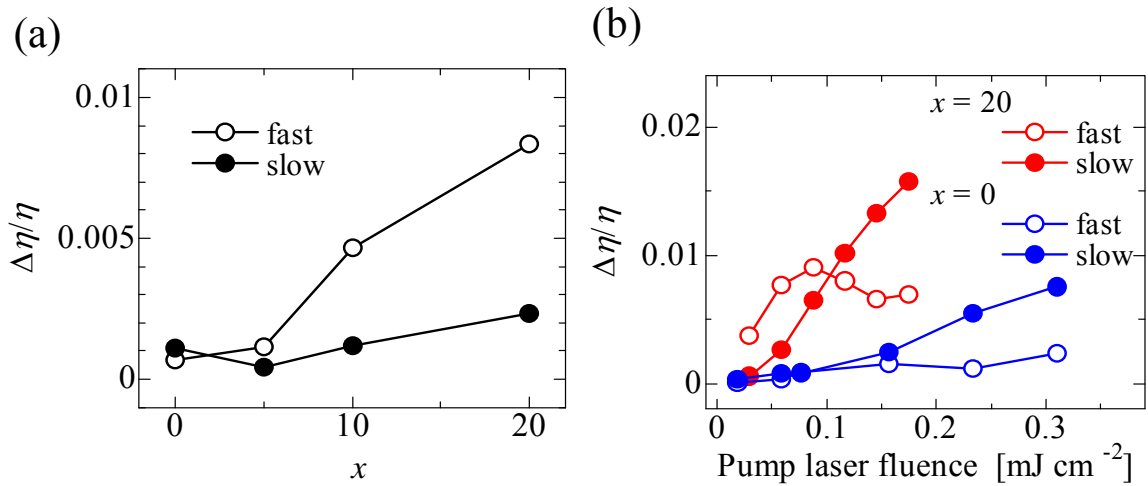

**Figure 7.** Fitted values of the fast- (open circles) and slow- (closed circles) recovering components for (a) $(Ag_x, Co):TiO_2/TiO_2$ films measured under a pump laser fluence of 0.06 mJ cm$^{-2}$ and (b) $(Ag_x, Co):TiO_2/TiO_2$ films for $x = 0$ (blue) and $x = 20$ (red) measured under various pump laser fluences.

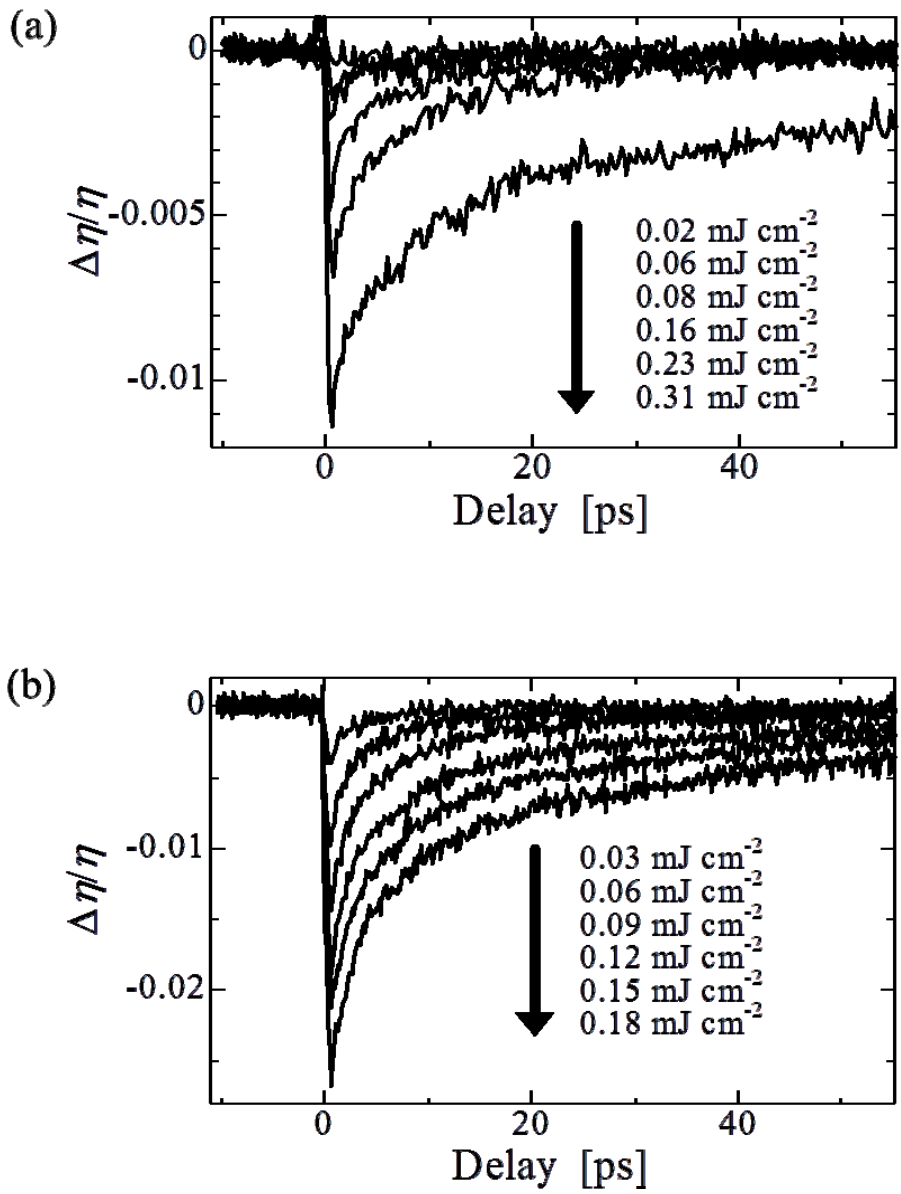

**Figure 8.** $\Delta\eta/\eta$ of (a) $x = 0$ and (b) $x = 20$ films.